\documentclass[11pt]{article}
\usepackage[footnotesize]{caption}
\usepackage[verbose,a4paper,tmargin=2.4cm,bmargin=2.4cm,lmargin=2.4cm,rmargin=2.4cm]{geometry}
\usepackage{times}
\usepackage{fullpage}
\usepackage{amsmath}
\usepackage{amsfonts}
\usepackage{amscd}
\usepackage{epsfig}
\usepackage{natbib}
\usepackage{setspace} 
\usepackage[displaymath]{lineno}

\newcommand{\mb}[1]{\mathbf{#1}}
\newcommand{\mr}[1]{\mathrm{#1}}
\newcommand{\ddt}[1]{\frac{d{#1}}{dt}}

\begin{document}

\title{\vspace{-1.75cm}Parameter Inference for an Individual Based Model of
  Chytridiomycosis in Frogs}

\author{Leah R.~Johnson, Cheryl J. Briggs\\
{\small Dept. of Ecology, Evolution, and Marine Biology}\\
{\small University of California Santa Barbara}\\
{\small Santa Barbara, CA 93106}\\
{\small Phone: (805) 893-2888}\\
{\small E-mail: {\tt lrjohnson@lifesci.ucsb.edu}, {\tt briggs@lifesci.ucsb.edu}}
}

\date{}

\maketitle

\runninglinenumbers
\doublespacing
\begin{center} {\bf Abstract} \end{center}
  Individual Based Models (IBMs) and Agent Based Models (ABMs) have
  become widely used tools to understand complex biological systems.
  However, general methods of parameter inference for IBMs are not
  available. In this paper we show that it is possible to address this
  problem with a traditional likelihood-based approach, using an
  example of an IBM developed to describe the spread of
  Chytridiomycosis in a population of frogs as a case study. We show
  that if the IBM satisfies certain criteria we can find the
  likelihood (or posterior) analytically, and use standard
  computational techniques, such as MCMC, for parameter inference.

\begin{center}
Parameter Inference; Bayesian Inference\\
\end{center}

\section{Introduction}

Individual or Agent Based Models (IBMs and ABMs, respectively) are widely used in ecology to understand how interactions between freely acting agents, or between components of a complex system, result in larger scale patterns and behavior \citep{grimm}. Typically, the foundation of an IBM consists of some kind of sub-model for ``individuals'', ranging from differential equations to heuristic rules. Multiple individuals are then allowed to interact via other rules within the virtual space.  Because they are fairly easy to formulate, are enormously flexible, and allow us to inherently incorporate individual variability, IBMs and ABMs have become increasingly popular for biological and ecological applications \citep{grimm:1999,huston:1988}. For instance, they have been used to describe  swarming or herding behaviors of animals \citep{gueron:1996,okubo:1986}, 
biofilm growth and formation \citep{kreft:1998,kreft:2001}, and fish population dynamics \citep{letcher:1996,madenjian:1991}, 
and white-nose syndrome in bats \citep{boyles:2009}. 

Although IBMs are useful for understanding the qualitative behavior of systems, quantitative analysis is less common. Although there are various reasons for this \citep{wiegand:2003}, a major factor is that model output is typically complex, and thus difficult to analyze and compare with data. Frequently the approach has been to use experimental or observational data on individuals within a population to obtain rough parameter estimates (such as birth or death rates) first, and then to use the IBM to explore the population level effects of various flavors of interactions given those parameters. However, in ecological settings, sometimes only system level observations are available, i.e., we cannot easily observe individuals uncoupled from each other or from the environment. Even if laboratory experiments on some parameters are available, these estimates may not translate to field conditions \citep{west:1996}. Thus, we would like to be able to use data from observational studies or other kinds of experiments that record individual level data in order to infer parameters in an IBM indirectly.

The recently developed approach of Pattern-Oriented Modeling (POM) \citep{grimm:2005,wiegand:2003} has been suggested as a way to standardize the analysis of IBMs, and facilitate indirect parameter inference. 
In practice, the procedure requires careful thinking about appropriate methods to compare model output and data. However, discussion of this, either inside or outside the POM framework, has been mostly ignored. Most methods for comparing model and data patterns for IBMs have been either qualitative or fairly ad-hoc, such as simple sensitivity analyses where parameters are perturbed over small ranges (e.g.~\cite{letcher:1996}). Although this may be sufficient for some applications, for others, such as the one we explore here, a more quantitative approach is required. In particular, we want to be able to perform parameter inference. 

For other mechanistic models, such as state-space models \citep{devalpine:2003}, methodologies for parameter inference via likelihood-based statistics are well-developed \citep{clark}. However, these tools have not been utilized for IBMs. In this paper we show that it is possible to derive a likelihood for classes of IBMs that have certain properties. We use a model of the dynamics of a fungal pathogen, {\em   Batrachochytrium dendrobatidis} (Bd), on and between frogs as a motivating example. First we introduce our case study and the IBM for which we would like to infer parameters. Next we discuss the derivation of the the likelihood, and show that we are able to infer parameters from simulated data using Markov chain Monte Carlo (MCMC). Finally we discuss under what circumstances practitioners should consider the approach: when it is likely to be possible to derive a likelihood in general, and how it can be practically extended for other models and data.

\section{Model}
Chytridiomycosis, a fungal disease of amphibians, has been implicated in  population declines and extinctions in amphibian populations worldwide \citep{berger:1998,skerratt:2007}. Efforts to understand the pathogenesis and spread of the disease have been wide ranging, including intensive observation of field populations \citep{lips:2006, murray:2009, vredenburg:2010}, 
laboratory experiments \citep{carey:2006,woodhams:2008}, and mathematical and statistical modeling \citep{briggs:2005,briggs:2010,mitchell:2008,ron:2005}.
However, many aspects of the disease, such as infection rates (both between and within frogs) or reproductive rates of the fungus, are very difficult to measure directly, especially in natural populations (e.g. \citep{rachowicz:2007}).  These factors are crucial for understanding why some populations are heavily affected by the disease \citep{lips:2006, vredenburg:2010}, while others are not \citep{briggs:2010,retallick:2004}, and for designing effective control strategies for the disease \citep{rachowicz:2005}. 

\subsection{The dynamics of Chytridiomycosis in a
  population} \label{model}
\begin{figure}[h!]
\begin{center}
\includegraphics[scale=0.375, trim=0 80 0 100, clip=true]{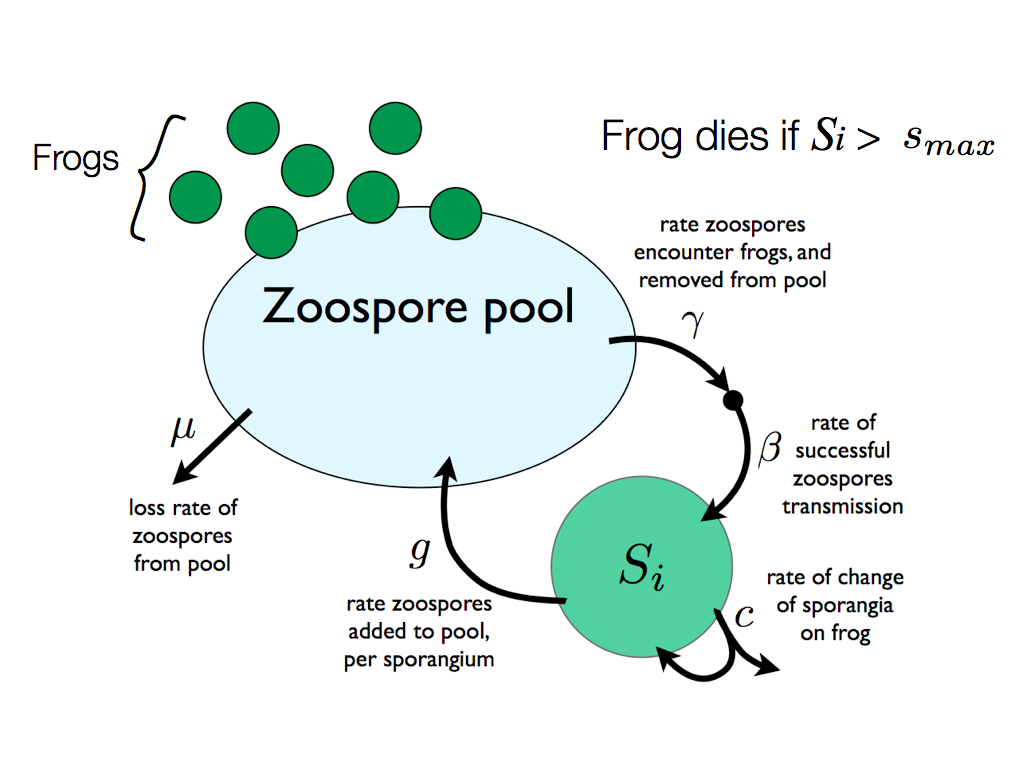} 
\end{center}
\caption[]{Schematic of deterministic model of the transmission of Bd in
  a frog population with a description of parameters included in the
model.}
\label{f:schematic}
\end{figure}

Bd is transmitted via an aquatic zoospore stage \citep{berger:2005,piotrowski:2001}, which infects keratinized cells on amphibian skin (or tadpole mouthparts; \cite{longcore:1999}). A proportion of the water-borne zoospores that encounter a cell successfully infect the cell and develop into a sporangia. These sporangia then release zoospores back onto the surface of the skin \citep{berger:2005}, where they can either reinfect the same individual, or enter the zoospore pool (the water containing zoospores, in which frogs live). Thus, infection within an individual and transmission between individuals are linked through a single process.  A series of IBMs (both deterministic and stochastic) describing this process were developed by \cite{briggs:2010}. Here we explore a version of this model for the spread of the disease within a single season, without stage (or age) structure or reproduction. The deterministic, continuous time dynamics (shown schematically in Figure \ref{f:schematic}, together with parameter definitions) are described by the following set of differential equations:

\vspace{-1.25cm}
\begin{align} \ddt{S_i}
  &= \begin{cases}
    \frac{\beta}{V} Z + cS_i & \text{ for   $S_i<s_{\mr{max}}$} \\
    0 \text{ and frog $i$ dies } & \text{ for $S_i>s_{\mr{max}}$} \label{eq:det:dS}
\end{cases}\\
\ddt{Z} &= \sum_{\text{all living frogs, $i$}} \left[ g S_i -
  \frac{\gamma}{V} Z \right] - \mu Z \label{eq:det:dZ}
\end{align} 

We are most interested in a discrete time stochastic version of this model \citep{briggs:2010}. To arrive at this model, time is first discretized into short steps of length $\Delta t$. Integrating Equations (\ref{eq:det:dS})-(\ref{eq:det:dZ}) over the time increment $t$ to $t+\Delta t$, holding the numbers of sporangia and zoospores fixed at the values at the beginning of the time step ($S_i(t)$ and $Z(t)$) approximates the mean of our stochastic process, which we choose to model as Poisson, since the numbers of sporangia and zoospores are counts.  Specifically, the number of sporangia $S_{i}(t+\Delta t)$ on individual $i$ (if it is still alive) at time $t+\Delta t$, given the sporangia and zoospore levels at the previous time ($S_i(t)$ and $Z(t)$), follows

\vspace{-1.25cm}
\begin{align} S_{i}(t+\Delta t) &\sim \mr{Pois}(\lambda_i(t+\Delta t)), \label{eq:S.distrib} \text{ where } \\ 
  \lambda_i (t+\Delta t) &\equiv E[ S_{i}(t+\Delta t) ]= S_{i}(t)
  e^{c \Delta t}+\frac{\beta Z(t)}{cV}\left( e^{c \Delta
    t}-1\right) \label{eq:S.mean}
\end{align}
Individuals die between $t$ and $\Delta t$ if the number of sporangia on the frog exceeds the lethal threshold, so that $S_i(t)>s_{\mr{max}}$, at which point all of the sporangia on the individual frog die as well (so that the sporangia load is defined to be zero).

The zoospore level at time $t + \Delta t$ depends on the zoospore level in the pool and the sum of sporangia across frogs, both at time $t$. The total zoospores in the pool at time $t+\Delta t$ has a Poisson distribution

\vspace{-1.25cm}
\begin{align}
Z(t+\Delta t) &\sim \mr{Pois}(\xi(t+\Delta T)), \label{eq:Z.distrib} \text{ where }\\
\xi (t+\Delta t)  &\equiv  E[Z(t+\Delta t)] = Z(t) e^{-c_3 \Delta
  t}+\frac{c_2}{c_3}\left( 1-e^{-c_3 \Delta t}\right)  \label{eq:Z.mean}
\end{align}
with $c_2 = \sum_{i=1}^{N} gS_{i}(t)$ and $c_3 = \mu + \frac{\gamma N}{V}$.
The model output for our likelihood based analysis is the sporangia loads on frogs and zoospore levels in the pool at each time step, and the times at which frogs die.

\subsection{Likelihood-based approach} \label{lik}

Many mechanistic models in ecology are formulated as deterministic differential equation models, which are frequently fit using non-statistical techniques, such as non-linear least squares \citep{hilborn_mangel,turchin:pop}. The use of Approximate Bayesian Computation has also been proposed for deterministic models \citep{toni:2009}.  Mechanistic models that are inherently stochastic or incorporate stochastic elements have also become popular to describe both populations and individuals. These types of models have been fit using statistical, typically likelihood based, methods. Some examples of these methods for biological models include maximum likelihood via iterated filtering \citep{ionides:2006} and flavors of particle filtering \citep{devalpine:2003,fujiwara:2005}. However, thus far these methods have not been used for parameter inference for IBMs.

In the classical or Bayesian statistical frameworks, the approach is to first find the analytic form of the likelihood, which is defined as the probability of observing the data under a parameterized model. Parameter inference, given data, then proceeds by finding maximum likelihood estimators (MLEs) or by sampling from the posterior distribution under particular priors. Each of the two approaches has its benefits. We find the Bayesian posterior summaries of uncertainty to be more intuitive compared to the classical approach of reporting confidence intervals, and so this is the approach we present here. MLEs and confidence bands could also be obtained using alternative computational approaches.

To infer the posterior probabilities of a vector of parameters $\theta=[\gamma, \mu, \beta, c, g, s_{\mr{max}}]$, we need to compute (and evaluate) the likelihood $\mr{Pr}((\mb{S}, \mb{A}, Z)_{1:T_{\mr{max}}} | \theta) $ where $(\mb{S}, \mb{A}, Z)_{1:T_{\mr{max}}}$ is the full data available on $N$ frogs over from time 1 to $T_{\mr{max}}$. For simplicity, in the following derivation we assume the data are observed without error, but with stochasticity inherent in the dynamics. Specifically, $\mb{S}$ is an $N \times T_{\mr{max}}$ matrix containing the sporangia counts on all individuals; $\mb{A}$ is an $N \times T_{\mr{max}}$ matrix that indicates whether individuals are alive (1) or dead (0); and $Z$ is a vector of length $T_{\mr{max}}$ which records the counts in the zoospore pool, at each time step.  $\mb{S}$ and $\mb{A}$ could also be viewed as a collection of $N$ vectors with length $\leq T_{\mr{max}}$, one for each frog. For the calculations below, we set $\Delta t = 1$ day. The derivation of the likelihood relies primarily on two conditions: the Markov property of the system and conditional independence of observations on individual frogs and the zoospore pool at a given time. By conditional probability, the likelihood can
be re-written as:

\vspace{-1.25cm}
\begin{align}
  L \equiv \mr{Pr}(&(\mb{S}, \mb{A}, Z)_{1:T_{\mr{max}}} | \theta) \\
  & = \mr{Pr}((\mb{S},\mb{A}, Z)_{T_{\mr{max}}} | (\mb{S}, \mb{A},
  Z)_{1:T_{\mr{max}}-1}, \theta)
  \times  \mr{Pr}((\mb{S}, \mb{A}, Z)_{1:T_{\mr{max}}-1} | \theta) \nonumber \\
  & = \mr{Pr}((\mb{S}, \mb{A}, Z)_{T_{\mr{max}}} | (\mb{S}, \mb{A}, Z)_{1:T_{\mr{max}}-1}, \theta) \nonumber\\
  & \hspace{0.5cm} \times \mr{Pr}((\mb{S}, \mb{A}, Z)_{T_{\mr{max}}-1} |
  (\mb{S}, \mb{A}, Z)_{1:T_{\mr{max}}-2}, \theta)
  \times \mr{Pr}((\mb{S}, \mb{A}, Z)_{1:T_{\mr{max}}-2} | \theta)\nonumber \\
  & \hspace{1cm} \vdots \nonumber \\
  & = \prod_{t=2}^{T_{\mr{max}}} \mr{Pr}((\mb{S}, \mb{A}, Z)_{t} |
  (\mb{S},\mb{A}, Z)_{1:t-1}, \theta).
\end{align}
In our case the new zoospore levels in the pool and sporangia loads on
the individual frogs depend only on the levels and loads at the previous
time step. In other words, the process exhibits a Markov property,
such that $ \mr{Pr}((\mb{S}, \mb{A}, Z)_{t} | (\mb{S},\mb{A},
Z)_{1:t-1}, \theta) = \mr{Pr}((\mb{S}, \mb{A}, Z)_{t} |
(\mb{S},\mb{A}, Z)_{t-1}, \theta)$, and so the likelihood can be
re-written as

\vspace{-1.25cm}
\begin{align*}
  L = \prod_{t=2}^{T_{\mr{max}}} \mr{Pr}((\mb{S}, \mb{A}, Z)_{t} |
  (\mb{S},\mb{A}, Z)_{t-1}, \theta)\nonumber
\end{align*}
If we look at the dynamic of the process over a single time step
(Equations (\ref{eq:S.distrib}) - (\ref{eq:Z.mean}) ), we can see
that, given the information about the state of the system at time
$t-1$, the zoospore load in the pool at time $t$ is independent of the
sporangia loads observed on frogs at time $t$. That is

\vspace{-1.25cm}
\begin{align*}
  L = \prod_{t=2}^{T_{\mr{max}}} \mr{Pr}((\mb{S}, \mb{A})_{t} |
  (\mb{S},\mb{A}, Z)_{t-1}, \theta) \times \mr{Pr}((Z)_{t} |
  (\mb{S},\mb{A}, Z)_{t-1}, \theta).
\end{align*}
Moreover, the observed levels on each frog are conditionally
independent of each other. In other words, given the state of the
system and parameters at time $t-1$, $(\mb{S},\mb{A})_t$
and $Z_t$ are conditionally independent, as are any pairs
$(S,A)_{i,t}$ and $(S,A)_{j,t}$ where $i \neq j$. Furthermore, the
state of individual $i$ at time $t$ only depends on the parameters,
its own state, and the reservoir at the previous time step
(i.e. $(S,A)_{i,t}$ is independent of $(S,A)_{j,t-1}$ for $i \neq j$)
so we can write the likelihood as

\vspace{-1.25cm}
\begin{align}
  L = \prod_{t=2}^{T_{\mr{max}}} \left(\prod_{i=1}^{N} \mr{Pr}((S, A)_{i,t}
    | (S, A)_{i,t-1}, Z_{t-1}, \theta) \right) \times \mr{Pr}((Z)_{t}
  | (\mb{S},\mb{A}, Z)_{t-1}, \theta). \label{eq:lik.full1}
\end{align}
The second term in the above product is the Poisson distribution
described in Equation (\ref{eq:Z.distrib}) with mean given by Equation
(\ref{eq:Z.mean}). Then the likelihood can be written as

\vspace{-1.25cm}
\begin{align}
  L = \prod_{t=2}^{T_{\mr{max}}} \left(\prod_{i=1}^{N} \mr{Pr}((S, A)_{i,t}
    | (S, A)_{i,t-1}, Z_{t-1} , \theta) \right)\times
  \prod_{t=2}^{T_{\mr{max}}} \frac{\xi(t)^{Z(t)}\exp{(-\xi(t))}}{Z(t)!}.
\end{align}
We can reverse the ordering of the products over $i$ and $t$ in the
first term of the above expression, and write things more compactly as

\vspace{-1.25cm}
\begin{align}
  L & = \prod_{i=1}^{N} \left(\prod_{t=2}^{T_{\mr{max}}} \mathcal{D}_{i,t}
  \right) \times \prod_{t=2}^{T_{\mr{max}}}
  \frac{\xi(t)^{Z(t)}\exp{(-\xi(t))}}{Z(t)!}, \label{eq:lik.full}
\end{align}
where

\vspace{-1.25cm}
\begin{equation}
  \mathcal{D}_{i,t} = \mr{Pr}((S, A)_{i,t}  | (S, A)_{i,t-1}, Z_{t-1} , \theta)
\end{equation}
is the probability of observing the data on the $i^{\mr{th}}$ frog at
time $t$ conditional on the data at the previous time step and the
parameters. By conditional probability, we can write this as

\vspace{-1.25cm}
\begin{align}
  \mathcal{D}_{i,t} = \mr{Pr}(S_{i, t} | A_{i,t}, (S, A)_{i,t-1},
  Z_{t-1} , \theta) \times \mr{Pr}(A_{i,t}| (S, A)_{i,t-1}, Z_{t-1} ,
  \theta).
  \label{eq:lik.frog1}
\end{align}
The probability that the individual is in state $A_{i,t}=\{0,1\}$ is
distributed as a Bernoulli random variable, conditional upon the
parameters, and all the data at time $t-1$,

\vspace{-1.25cm}
\begin{align}
  \mr{Pr}(A_{i,t}| (S, A)_{i,t-1}, Z_{t-1} , \theta) =
  p_{i,t}^{A_{i,t}}(1-p_{i,t})^{1-A_{i,t}}, \label{eq:bern}
\end{align}
where the probability of a success, $p_{i,t}$, defined as the frog being alive at
time $t$ ($A_{i,t}=1$), is the probability that the number
of sporangia on the frog will less than $s_{\mr{max}}$, i.e.,

\vspace{-1.25cm}
\begin{align}
  p_{i,t}=F_{s_{\mr{max}}}(\lambda_i(t))\label{eq:p.surv}
\end{align}
where $F_{x}(q)$ denotes the cumulative distribution function up to
$x$ for the Poisson distribution with parameter $q$. Conditional on
$A_{i,t}$, then, the probability of observing the sporangia load on
the frog is given by

\vspace{-.75cm}
\begin{equation}
  \mr{Pr}(S_{i, t} |  A_{i,t},  (S, A)_{i,t-1}, Z_{t-1} , \theta)= 
\begin{cases}
  \mr{Pois}(\lambda_i(t)) & \mr{if \hspace{0.2cm} } A_{i,t}=1,\\
  \delta_{(S_{i,t},0)} & \mr{if \hspace{0.2cm} } A_{i,t}= 0.
\end{cases} \label{eq:s.step}
\end{equation}
where $\delta_{(S_{i,t},0)}$ is the Kronecker delta function. Note
also that, for completeness, we can specify that $\mr{Pr}(A_{i,t}=0|
A_{i, t-1}=0) =1$, although the data we observe ensure this. Let $T_i$
be the last time that the frog is observed alive. Using this together
with Equations \ref{eq:bern} and \ref{eq:s.step}, we can then re-write
the expression for $\mathcal{D}_{i,t}$ as

\vspace{-.25cm}
\begin{equation}
\mathcal{D}_{i,t} = 
\begin{cases}
  \left(\frac{\lambda_i(t)^{S_{i,t}}\exp{(-\lambda_i(t))}}{S_{i,t}!}
  \right) \times p_{i,t} & \text{ if } t \leq T_i \\
  1-p_{i,t} & \text{ if } t = T_i+1 \\
  1 & \text{otherwise}
\end{cases} \label{eq:D}
\end{equation}
Incorporating the expression for $\mathcal{D}_{i,t}$ into Equation
\ref{eq:lik.full}, we arrive at an expression for the full likelihood:

\vspace{-1.25cm}
\begin{align}
  \mr{Pr}((\mb{S}, \mb{A}, Z)_{1:T_{\mr{max}}} | \theta) & =
  \prod_{i=1}^{N}  \left(\prod_{t=2}^{T_{\mr{max}}} \mathcal{D}_{i,t}
  \right) \times \prod_{t=2}^{T_{\mr{max}}}
  \frac{\xi(t)^{Z(t)}\exp{(-\xi(t))}}{Z(t)!} \nonumber \\
  &= \prod_{i=1}^{N} \left(\prod_{t=2}^{T_i}
    \left(\frac{\lambda_i(t)^{S_{i,t}}\exp{(-\lambda_i(t))}}{S_{i,t}!}
    \right) \times p_{i,t} \right) \times (1-p_{i,T_i+1})\nonumber \\
  & \hspace{0.4cm} \times \prod_{t=2}^{T_{\mr{max}}}
  \frac{\xi(t)^{Z_t}\exp{(-\xi(t))}}{Z_t!}.
\end{align}

\subsection{Simulation Experiment} \label{methods}

\begin{figure}[h!]
\begin{center}
\includegraphics[scale=0.5, trim=0 40 0 0]{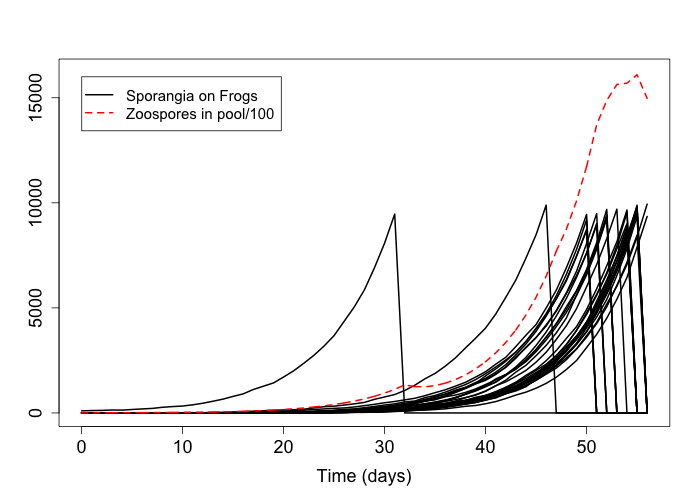}
\end{center}
\caption[]{One realization of the simulation model shown in Figure \ref{f:schematic}. Shown are sporangia loads on frogs all frogs (N=30, solid black lines), and the zoospore
  level in the pool (red dotted line) over the course of the virtual experiment. Each
  solid line represents the sporangia load on an individual frog. When
  the sporangia load reaches a maximum (here 10000), the frog dies, and the
 sporangia load on that frog immediately drops to zero.}\label{f:realization}
\end{figure}

Once the likelihood has been derived, various approaches could be used for parameter inference. As an example of how this can proceed, we seek to infer parameters using a single stochastic realization of the model described in Section \ref{model} as data. The simulation consists of 30 individuals in a single pool of unit volume. Initially, the pool is free of zoospores, and all individuals, save one, are uninfected. The infected individual that acts as the source of the infection has a load of 100 sporangia. Parameter values used in the simulation are given in Table \ref{tb:priors}. This data realization is shown in Figure \ref{f:realization}.

We attempt to infer parameter values for three cases: data without noise, data with normal noise, and data with lognormal noise. In the first case we assume that we can monitor both the level of the infection on individual frogs as well as zoospore levels within the pool (or reservoir) without any observation error (i.e.~all stochasticity is due to process error). In the second and third cases we add a small amount of observational noise to the data to see if the inferential methods are robust. In our second case, we consider additive noise so that the observed values, $\tilde{S_i},\tilde{Z}$ are given by $\tilde{S_i} = S_i + \epsilon_1$ and $\tilde{Z} = Z + \xi_1$ where $\epsilon_1, \xi_1 \sim \mr{N}(0,10)$. In the third case we assume multiplicative noise, so that $\tilde{S_i} = S_i e^{\epsilon_2}$ and $\tilde{Z} = Z e^{\xi_2}$ where $\epsilon_2, \xi_2 \sim \mr{N}(0,0.01)$. Throughout, we assume that we do not have false positives, {\em i.e.}, we correctly identify uninfected individuals as having zero fungal load ($S_i=0$).

The likelihood is a sufficiently complicated function of the data and parameters that analytical methods are not feasible. Thus we turn to numerical methods, specifically Markov chain Monte Carlo (MCMC) methods in the Bayesian context. We specify weak priors on all parameters (see Table \ref{tb:priors}) and sample from the posterior distribution of the parameters using a variation of the random walk Metropolis-Hastings (MH) algorithm. A short description of the algorithm used and its implementation can be found Appendix \ref{ap:MH}. See \cite{clark} for further details of these sampling approaches in the context of ecological problems. For all cases, samples of 20000 draws from the posterior distribution were collected.

\section{Results}

\begin{figure}[h!]
\begin{center}
\includegraphics[scale=0.425, trim=0 30 0 0]{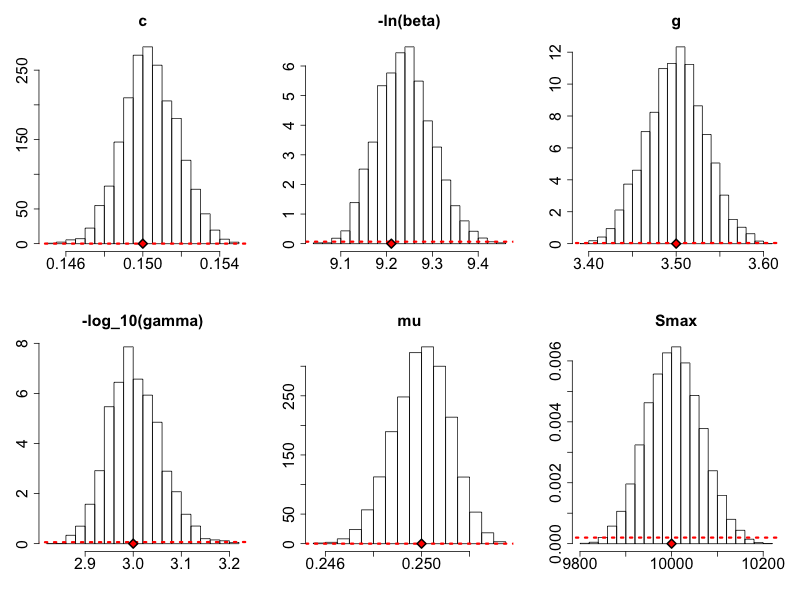}
\end{center}
\caption[]{Histograms of marginal posterior densities of the model
  parameters estimated from data without observation error. The
  ``true'' value of the parameter is indicated with a red
  diamond. Prior distributions are shown as dotted red
  lines.}\label{f:post1}
\end{figure}

We begin with inference for the system without observational error. Figure \ref{f:pairs1} shows the samples from the joint posterior distributions plotted for pairs of parameters in this case. The marginal posterior (and prior) distributions of the parameters are shown in Figure \ref{f:post1}. As is clear in the figure, we were able to get very good posterior distributions for all of the parameters, in that the posterior 95\% credible interval includes the ``real'' parameter values used to generate the data and the posterior has low variance. Additionally, we can see that the prior distributions for the parameters in the regions of high posterior probability were very uninformative (i.e.~the priors are low and flat in these regions). 

\begin{figure}[h!]
\begin{center}
\includegraphics[scale=0.425, trim=0 0 0 0]{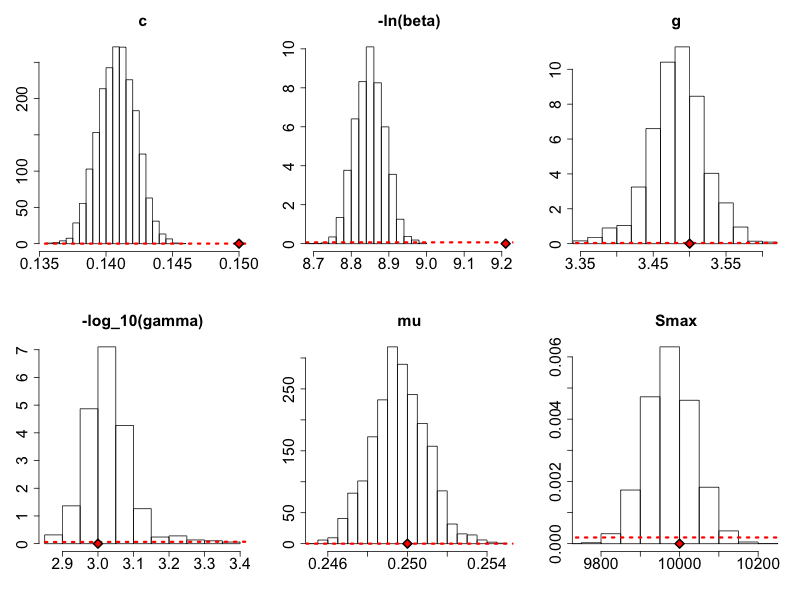} \\
\includegraphics[scale=0.425, trim=0 30 0 0]{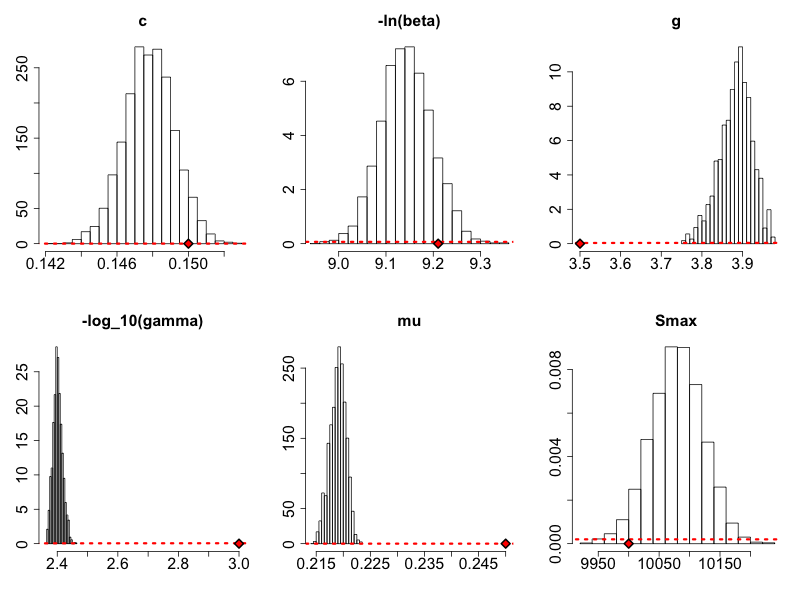}
\end{center}
\caption[]{Samples from the posterior distributions of parameters when
  data observed sampled with a small amount of (top two rows) normally
  distributed noise (N(0,10)) and (bottom two rows) Log-Normally
  distributed noise (logN(0,0.01)) The ``true'' value of the parameter
  is indicated with a red diamond. Prior distributions are shown as
  dotted red lines. See Section \ref{methods} for
  details.}\label{f:hist:n1}
\end{figure}

Next we examine the two cases where additional observation error is added to the data, but the model is not extended to include an observational model. Marginal posterior distributions of the parameters for both of these cases are shown in Figure \ref{f:hist:n1}. Notice that although the posterior distributions are quite different from the prior (i.e., the data are informative for the parameters) only some are infered with similar accuracy to the case without noise. In particular, $s_{\mr{max}}$ is reasonably accurately inferred in both cases. However, as there is additional variability in the noisy data compared to what would be typical in output from the model with ``correct'' parameters, some parameter estimates are biased.  Which parameter estimates are biased depends upon the type of noise. For the case with additive, normal error, estimates for $c$ and $\beta$ (mostly effecting the dynamics of sporangia on frogs) are biased, whereas for the multiplicative log-normal error, $g$, $\gamma$, and $\mu$ (mostly effecting the dynamics of the zoospore pool) are biased.

This pattern of bias is related to how the error enters into the model, which effects the relative size of the noise compared to the signal for sporangia and zoospore counts. The zoospore levels in the pool are typically more than two orders of magnitude higher than the sporangia levels on an individual frog (see Figure \ref{f:realization}). Thus for the additive noise case, the relative errors in the sporangia counts is larger than for the zoospore levels. Since the observational noise is smaller than the process error the posterior estimates for the parameters determining the zoospore dynamics are largely unchanged. However, for the log-normal noise, this is not true. Instead, as the value of the counts increases, the absolute error increases as well, similarly to how the process stochasticity increases with the mean. For the zoospore load in the pool, the noise is able to skew the parameter estimates that are primarily informed by this time series. Although we might expect similar results for the sporangia on frogs, since there are many trajectories being observed simultaneously, the estimates of parameters related to sporangia dynamics are less likely to be biased in any particular direction. Observe, that the overall magnitude of the multiplicative noise is larger than in the additive case, and thus the accuracy of the estimates for all parameters is poorer.

\section{Discussion}

Mathematical models in ecology and biology play many roles, from providing short term prediction, to acting as tools for furthering our general understanding of biological mechanisms. Frequently, as has been the case with most IBMs, mechanistic models have been constrained to be used for {\em   qualitative} understanding as methods for fitting them to data have not always been available. However, as we have shown here, it may be possible to describe an IBM using a likelihood based approach and perform indirect parameter inference from data for quantitative predictions.

Finding an appropriate likelihood for an IBM requires an additional investment above and beyond the current practice of building a simulation model. Thus, the choice to take this approach necessarily depends upon the goal of the modeling exercise. If the primary objective is to understand the range of system behaviors for different types of individual actions and interactions then this approach may be overkill. However, if there is a real desire to be able to perform parameter inference from data, especially in cases where knowledge of the uncertainty in the estimates is desired, the investment is likely to yield more satisfying results than other approaches currently being used.

In the system explored here, there are a number of reasons why we particularly want to be able to estimate parameters, instead of just understanding the behavior of the model or making qualitative predictions. For instance, we want to better understand the biological differences between strains of the fungus, between species of frogs affected by the fungus, the effects of temperature variation on the spread of the epidemic, etc. These differences manifest themselves as variation in parameter values, which are now estimable. Knowing both parameter estimates, and the uncertainty in these estimates, allows the design of effective intervention strategies \citep{elderd:2006,merl:2009}. The approach shown here has the additional feature that by explicitly taking into account interactions between individuals we can utilize all available data for parameter inference while reducing concerns of pseudo-replication.

The general assumptions that allow us to find the likelihood for the IBM presented here would be applicable to many, though not all, other IBMs. Certain assumptions in many IBMs could make this approach more difficult.  For instance, models where individuals exhibit behavioral ``rules of thumb'' or models that include spatially explicit stochastic movement and interactions between individuals, may be considerably more difficult to formulate in this framework. However, even in these cases it may be possible to find an appropriate likelihood model for the data (for instance see \citet{patterson:2008}), or to consider simpler or non-heuristic version of the model that is amenable to a likelihood based analysis. For the IBM described in this paper the most important simplifying assumptions are: 1) that transitions are described by parametric distributions; 2) that individuals are conditionally independent at the current time step, so that their current state only depends on the states of other individuals at previous times; 3) the system exhibits a Markov property (the state at time $t$ only depends on the state at time $t-1$). In fact, when the Markov property holds, this type of IBM could be viewed as a type of statistical state-space model (SSM) \citep{buckland:2004,buckland:2007,newman:2006}, and thus one could use many of the computational tools that have already been developed by the statistical community for analysis of SSMs \citep{harvey:2004,newman:2009}. Because of this, we envision straightforward parametric inference when various extensions are added to the model presented here. For instance, observational uncertainty, hidden states, hierarchical structures (all three of which are standard in SSMs), and more biological complexities, are easily incorporated.

Of the extensions mentioned above, the addition of an observation model is likely to be the most important, since, as we showed, ignoring this source of stochasticity can lead to bias in parameter estimates. The addition of an observation model is conceptually simple when the model is treated as an SSM with an unobserved state process. However, when portions of the state are hidden, either because of the inclusion of an observation model, or because there is no data gathered for a portion of the process, inference may be more difficult. This may be because there is not enough data to infer the hidden state, or because the computational routines necessary for inference are more involved. Additionally, the tools best suited for inference in these cases are frequently less well known to ecology community. For instance, MCMC and Sequential Importance Sampling (SIS), both standard methods in statistics, have been used for Bayesian inference for SSMs \citep{newman:2009}, and these approaches are slowly being introduced in the ecological literature.

Although some types of extensions would typically require different computational techniques to perform the analysis, not all extensions have this limitation. Since IBMs are typically designed to explore how small differences between individuals propagate through the system, the addition of hierarchical modeling \citep{clark:gelfand,mcclintock:2010}, 
for instance to explicitly include variation in parameter values between populations or individuals, would be a natural extension that is still tractable with simple methods.  Additionally, various other biological details, such as age or state structure, could be incorporated into the modeling framework fairly easily, as long as the Markov property, etc., are maintained. Thus, although the computational burden of likelihood-based inference for this approach maybe more involved than simply performing forward simulations of the IBM, this cost can be more than offset by the ability perform robust quantitative analysis and parameter inference.

\section{Acknowledgments}
We would like to thank Robert Gramacy and Roger Nisbet for helpful comments on earlier drafts.  This work was supported by NSF grant EF-0723563 to CJB.

\appendix

\section{Sampling from a Distribution via the Metropolis Hastings or
  Metropolis-within-Gibbs algorithms}\label{ap:MH}
The Metropolis-Hastings (MH) algorithm is a type of Markov chain Monte
Carlo (MCMC) simulation commonly used to sample from distributions,
particularly posterior distributions in Bayesian statistics, when
other sampling methods are not available. Details on MCMC and the
various sampling methods discussed below can be found in
\citet{clark}. Here we give a very brief outline of the sampling
methods utilized in this paper. In the most basic sense, the MH
algorithm begins at some parameter setting, proposes new values with a
user-specified probability related to the location of the parameters relative to the
original sample, evaluates the posterior probability density of the
new sample, then ``accepts'' the new sample probabilistically, where
this probability is proportional to the posterior of the last and
proposed samples and the probability of the proposed samples. In particular, the acceptance ratio for the MH
algorithm is defined to be

\vspace{-1cm}
\begin{equation*}
A(\theta^l|\theta^*)=\frac{p(\theta^*)/J(\theta^*|\theta^l)}{p(\theta^l)/J(\theta^l|\theta^*)}
\end{equation*}
where $\theta^l$ is the last accepted draw of the parameters,
$\theta^*$ is the proposed parameter set, $p(\cdot)$ denotes the
distribution we are trying to sample from, in our case the posterior
distribution, and $J(\theta^2|\theta^1)$ the jump (or proposal)
distribution, which gives the probability of proposing parameters
$\theta^2$ given that the current parameters are $\theta^1$.  More
specifically, the algorithm is as follows:
\begin{itemize}
\item Choose an initial setting of the parameter vector $\theta^{(0)}$
\item For $\tau$ in 0 to $\tau_{\mr{max}}$
  \begin{enumerate}
  \item Set $\theta^{(\tau+1)}=\theta^{(\tau)}$
  \item propose a new value of the parameters 
      $\theta^*$ 
    \item evaluate the MH acceptance ratio $A(\theta^l|\theta^*)$ (defined above)
    \item choose $u\sim \mr{Unif}(0,1)$
    \item accept the proposed parameter if $u<\mr{min}\{A,1\}$, i.e.,
      set $\theta^{(\tau+1)} = \theta^*$ (otherwise
      $\theta^{(\tau+1)}$ remains fixed at $\theta^{(\tau)}$)
  \end{enumerate}
\end{itemize}

The Metropolis-within-Gibbs algorithm is a variant of the MH algorithm
which allows one to update parameters individually, instead of all at
once. The algorithm proceeds as follows:
\begin{itemize}
\item Choose an initial setting of the parameter vector $\theta^{(0)}$
\item For $\tau$ in 0 to $\tau_{\mr{max}}$
  \begin{itemize}
  \item Set $\theta^{(\tau+1)}=\theta^{(\tau)}$
  \item For $i$ in 1 to $N$:
    \begin{enumerate}
    \item propose a new value of the $i^{\mr{th}}$ parameter 
      $\theta_i^*$, so that $\theta^* = \theta_{-i} \cup \theta_i^*$.
    \item evaluate the MH acceptance ratio $A(\theta^l|\theta^*)$ (defined above)
    \item choose $u\sim \mr{Unif}(0,1)$
    \item accept the proposed parameter if $u<\mr{min}\{A,1\}$, i.e.,
      set $\theta_i^{(\tau+1)} = \theta_i^*$
    \end{enumerate} 
  \end{itemize}
\end{itemize}
Additionally, instead of proposing parameters only one at a time,
parameters can be chosen in blocks of multiple parameters at a
time. This can be especially useful when parameters are correlated and
a joint jump distribution is used, as this can improve mixing in the
chain.

\begin{figure}[h!]
\begin{center}
\includegraphics[scale=0.75, trim=0 30 0 0, clip=true]{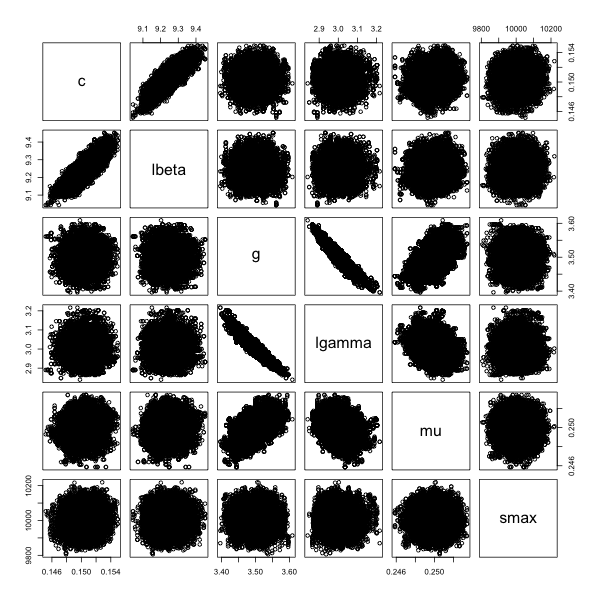}
\end{center}
\caption[]{Samples from the joint posterior distribution of parameters
  obtained via MCMC, plotted for all pairs of parameters. The
  distribution is estimated from data without observation
  error. } \label{f:pairs1}
\end{figure}

For our model, we choose to use the MwG algorithm. The likelihood used
was that derived in Section \ref{ap:LK}. Proposal distributions for
the parameters were normal, centered at the previously accepted
parameter value (or multi-variate normal, for jointly proposed
parameters), with variances (and covariances) that were tuned in
preliminary runs so that an appropriate balance of acceptance and
mixing was achieved.  Figure \ref{f:pairs1} shows the samples from the
joint posterior distributions plotted for pairs of parameters obtained
using this method. Although most pairs of parameters are fairly
independent, the exceptions are the strong correlations between $c$
and $\beta$ (or in the plot $\ln{\beta}$) and between $g$ and $\gamma$
(or $\log_{10}{\gamma}$). Thus we chose to jointly propose $g$ with
$\log_{10}(\gamma)$), which significantly improved mixing.  We
observed that even when the Markov chain was initialized away from the
true values the sampler moved fairly quickly into the appropriate
portion of parameter space. For the two cases with error, a burn-in of
approximately 1000 samples seemed to be sufficient.

\footnotesize
\bibliographystyle{plain}%

\begin{table}[h!]
\begin{center}
\begin{tabular}{| c c c c | }
  \hline
  parameter & value & prior & posterior 95\% C.I. \\
  \hline
  $-\log_{10}(\gamma)$ & 3 & $\mr{Gamma}(3,3)$ & (2.90 -- 3.11) \\
  $\mu$ &  0.25 &$\mr{Gamma}(3,2)$ & (0.248 -- 0.252)\\
  $-\ln(\beta) $ & 9.21 & $\mr{Gamma}(3,4)$ & (9.12 -- 9.36)\\
  $c$ & 0.15 & $\mr{N}(0,5)$ & (0.148 -- 0.153)\\
  $g$ &  3.5 & $\mr{Gamma}(3,4)$ & (3.44 -- 3.56)\\
  $s_{\mr{max}}$ & 10000 & $\mr{N}(10000,2000)$ & (9890 -- 10120)\\
  \hline
\end{tabular} 
\caption[]{Parameter values used for the ``data'' simulation, and
corresponding prior distributions and 95\% central posterior
  credible intervals estimated from data without observation
  error.  \label{tb:priors}}
\end{center}
\end{table}

\end{document}